\documentclass[aps,amsmath,amssymb,prl,twocolumn,showpacs]{revtex4}
\usepackage{bm}
\usepackage{epsfig}

\begin{document}

\preprint{draft}

\title{Casimir Forces: An Exact Approach for Periodically Deformed Objects}

\author{Thorsten Emig}

\affiliation{Institut f\"ur Theoretische Physik, Universit\"at zu
K\"oln, Z\"ulpicher Stra\ss e 77, D-50937 K\"oln, Germany}

\date{\today}

\begin{abstract}
  A novel approach for calculating Casimir forces between periodically
  deformed objects is developed. This approach allows, for the first
  time, a rigorous non-perturbative treatment of the Casimir effect
  for disconnected objects beyond Casimir's original two-plate
  configuration. The approach takes into account the collective nature
  of fluctuation induced forces, going beyond the commonly used
  pairwise summation of two-body van der Waals forces. As an
  application of the method, we exactly calculate the Casimir force
  due to scalar field fluctuations between a flat and a rectangular
  corrugated plate. In the latter case, the force is found to be
  always attractive.
\end{abstract}


\maketitle

Casimir forces between electrically neutral solids are a macroscopic
consequence of material dependent changes in the zero-point vacuum
fluctuations of the electromagnetic field
\cite{Casimir48,Milonni+Mostepanenko-book,Bordag-review}.  The Casimir
effect for two parallel metallic plates has been referred to as one of
the least intuitive consequences of quantum electrodynamics
\cite{Schwinger+78}.  However, this effect can be as well regarded as
a macroscopic manifestation of many-body van der Waals forces between
the particles that form the plates.  This point of view is supported
by Schwinger's approach which considers no vacuum fluctuations but
only the fields generated by the fluctuating dipoles themselves
\cite{Schwinger+78}.  In addition, fluctuation induced forces can be
observed in a plethora of other systems where quantum-, thermal or
disorder fluctuations are modified by external objects
\cite{Kardar+99}. Examples of recent interest are as diverse as the
thickening of Helium films near the superfluid transition
\cite{Israelachvili-book,Garcia+02}, forces between vortex matter in
anisotropic superconductors \cite{Buechler+00/Mukherji+97}, Casimir
energy densities in cosmological models \cite{Bytsenko+96}, and
interactions between proteins on biological membranes
\cite{Israelachvili-book}.

Initiated by the first quantitative verification of the electrodynamic
Casimir effect by Lamoreaux \cite{Lamoreaux97}, high precision
experiments have motivated a resurgence in the field of Casimir force
measurements in the last few years \cite{Mohideen+Chan+Bressi}. All
experiments confirm the theory for Casimir's original flat plate
geometry within a few percent of accuracy. Much less clear is the
situation for non-trivial geometries. If the separation between the
objects is large compared to characteristic wavelengths of the
material, e.g., the plasma wavelength, the force becomes a universal
function of the geometry with the energy scale set by $\hbar$.
However, the entropic origin and collective nature of the Casimir
force can cause this universal function to have a non-trivial and
unexpected dependence on the shape of the interacting objects. There
is little intuition even as to whether the interaction is attractive
or repulsive as demonstrated strikingly by the {\it positivity} of the
Casimir energy of a conducting spherical shell \cite{Boyer74}. In
particular, the latter result raises the important question if
repulsive forces can also emerge between disconnected objects, being
of potential importance for the behavior of microelectromechanical
systems due to the short-scale separations between their mobile parts
\cite{Serry+95}.  Recently, non-trivial shape dependencies have been
probed experimentally by tailoring the shape of the plates in
Casimir's geometry \cite{Roy+99,Chen+02}.

Calculations of Casimir forces beyond the simple situation of two
parallel flat plates are notoriously difficult due to the collective
nature of the many-body interaction. To date, this simple geometry
indeed appears to be the only case amongst disconnected macroscopic
objects for which an exact result is available.  This is due to the
lack of rigorous, non-perturbative methods for calculating the force
between deformed objects. The simplest and commonly used approximation
is the pairwise summation (PWS) of van der Waals forces
\cite{Bordag-review}.  However, Lifshitz's theory for dielectric
bodies (with flat surfaces) demonstrates that in general the
interaction cannot be obtained from such an approximation
\cite{Lifshitz56}.  In view of the fact that the PWS approximation can
even give the wrong sign for the Casimir interaction, e.g., for a
conducting cubic shell \cite{Barton01}, it is fair to conclude that
this approach is to some extent uncontrolled. For corrugated metal
plates, the failure of PWS for small corrugation lengths has been
shown by perturbation theory with respect to the deformation amplitude
\cite{Emig+01}. Another perturbative approach is based on a multiple
scattering expansion but has been applied only in the limit of large
separations between the objects \cite{Balian+78}.  However, with
regard to potential sign changes in the interaction and experimentally
realized separations, it is highly desirable to gather information
about the perturbatively non-accessible regime where the distance
between the objects is of the same order as the deformation amplitude.

In this Letter we will introduce an avenue along which the
shortcomings of the commonly used approximations can be bypassed for
objects with uniaxial shape modulations. To illustrate our new
approach, we consider the force between a rectangular corrugated and a
flat plate, a geometry which cannot be treated by perturbation theory
due to the edges in the surface profile. Before coming to this
specific case, let us outline the general approach. In uniaxial
geometries the electromagnetic field and thus also the Casimir force
can be represented as a sum of transversal magnetic (TM) and electric
(TE) wave contributions \cite{Emig+01}.  In order to simplify the
analysis, we will consider here only TM waves, which are governed by
the scalar field Euclidean action
\begin{equation}
S[\Phi]=\frac{1}{2}\int d^4 X (\nabla \Phi)^2
\end{equation}
after a Wick rotation to imaginary time. For TM waves we have to
impose Dirichlet boundary conditions, $\Phi_{|S}=0$ on the surfaces
$S$. TE waves can be treated analogously by imposing Neumann boundary
conditions with suppressed normal gradient of $\Phi$.  The geometry
consists of a flat surface in the $xy$-plane at $z=H$ and a
uniaxially corrugated surface given by $z=h(x)$ with periodic height
profile $h(x+\lambda)=h(x)$ of wavelength $\lambda$.  At zero
temperature, the Casimir force per surface area $F=-\partial {\cal
  E}/\partial H$ between the two objects corresponds to the change
with the objects distance in the ground state energy density ${\cal
  E}$ of the field $\Phi$. This zero temperature theory is applicable
as long as the de Broglie wavelength of photons, $\lambda_T=\hbar
c/k_B T$, is much larger than the mean separation between the objects.

Employing the path integral method to implement the boundary
conditions at the surfaces \cite{Li+91,Golestanian+97}, the energy
density is given by ${\cal E}=-\hbar c \ln {\cal Z}/AL$ with $L$ the
Euclidean system size along imaginary time and
\begin{equation}
{\cal Z}={\cal Z}_0^{-1} \int {\cal D} \Phi \prod_{\bf r} 
\delta[\Phi({\bf r},h(x))] \delta[\Phi({\bf r},H)] 
e^{-S[\Phi]/\hbar},
\end{equation}
with the partition function ${\cal Z}_0$ of empty space with no
objects, and ${\bf r}=(ict,x,y)\equiv (X_0,X_1,X_2)$.  The delta
functions can be rewritten as a functional integral over an auxiliary
field on each surface so that after integrating out $\Phi$, the new
action is quadratic with matrix kernel
\begin{equation}
M({\bf r},{\bf r}')=\left(\!\!
\begin{array}{cc}
G[{\bf r}-{\bf r}',h(x)\!-\!h(x')] & G[{\bf r}-{\bf r}',H\!-\!h(x')] \\
G[{\bf r}-{\bf r}',H\!-\!h(x)] & G[{\bf r}-{\bf r}',0]
\end{array}\!\!\right)
\end{equation}
and Green's function $G({\bf r},z)=({\bf r}^2+z^2)^{-1}/4\pi^2$
\cite{Hanke+02}. The Casimir force per surface area $A$ between the
objects is then given by
\begin{equation}
\label{force}
F=-\frac{\hbar c}{2AL} \partial_H \ln \det M
=-\frac{\hbar c}{2AL} {\rm Tr}\left( M^{-1} \partial_H M \right).
\end{equation}
This expression is suited for a non-perturbative approach since the
force must be finite, in contrast to the energy density ${\cal E}$
which contains without an explicit cutoff also infinite but $H$
independent contributions from the individual surfaces.

In the more conventional approaches, the zero-point energy is
calculated in terms of the allowed frequencies $\omega_n$ of the space
between the objects, using ${\cal E}=\frac{1}{2}\sum_n \hbar\omega_n$
and by applying subtraction schemes in order to obtain a regularized
energy which in turn yields the force. An advantage of our approach is
that the frequencies need not to be calculated explicitly but the
force is directly obtained in terms of the {\it free} space Green's
function without any necessity for regularization schemes.  Of course,
the functional inverse of the matrix kernel $M({\bf r},{\bf r}')$ in
Eq. (\ref{force}) is in general difficult to evaluate. In the
following, we will demonstrate how this problem can be tackled for
uniaxial geometries. Due to the symmetry of the geometry, it is useful
to consider the matrix kernel $M({\bf p},{\bf q})$ in momentum space.
The periodicity of the surface profile along the $X_1$ direction and
uniformity in the $X_0 X_2$ plane impose the general form
\begin{eqnarray}
\label{decomp}
M({\bf p},{\bf q})&=&\sum_{m=-\infty}^\infty  N_m(p_\perp,p_1) 
\, (2\pi)^3 \, \delta({\bf p}_\perp+{\bf q}_\perp) \nonumber \\
&&\times \, \delta(p_1+q_1+2\pi m/\lambda)
\end{eqnarray}
on the kernel. This Fourier decomposition defines the matrices
$N_m(p_\perp, p_1)$ with ${\bf p}_\perp=(p_0,p_2)$ and $p_\perp=|{\bf
  p}_\perp|$. Thus the matrix $M({\bf p},{\bf q})$ has non-vanishing
entries only along the diagonal and its periodically shifted
analogues. An important observation is that matrices of this structure
can be transformed to block-diagonal form by applying row and column
permutations.  Physically, the existence of such a transformation is
due to the non-mixing property of the kernel $M({\bf p},{\bf q})$ for
Fourier modes whose moments differ by non-integer multiples of
$2\pi/\lambda$.  For the time being, let us assume that the objects
have the extension $W$ along the $X_1$ direction, leading to a
discrete set of momenta $p_1$. Then the transformed matrix $M({\bf
  p},{\bf q})$ consists of block matrices $M_j$ along the diagonal
with $j=1,\ldots, N=W/\lambda-1$. Since this matrix form can be always
realized by an even number of permutations, we get
\begin{equation}
\label{factor}
\det M = {\prod}_{j=1}^N \det M_j.
\end{equation}
The block matrices $M_j$ can be expressed in terms of the matrices
$N_m$ of the decomposition in Eq. (\ref{decomp}). For discrete momenta
in $x$ direction, the matrices $M_j$ can be parametrized by integer
indices $k$, $l=-\infty,\ldots,\infty$, leading to the result
\begin{equation}
M_{j,kl}({\bf p}_\perp, {\bf q}_\perp)=(2\pi)^2\delta({\bf p}_\perp+
{\bf q}_\perp) B_{kl}(p_\perp,2\pi j/W)
\end{equation}
with the matrix $B_{kl}$ given by
\begin{equation}
\label{matrix-B}
B_{kl}(p_\perp,p_1)=N_{k-l}(p_\perp,p_1+2\pi l/\lambda).
\end{equation}
Moreover, from Eq. (\ref{factor}) we obtain easily $\partial_H (\ln
\det M)=\sum_{j=1}^N {\rm Tr} (M_j^{-1} \partial_H M_j)$. Thus the
separation dependent part of the ground state energy ${\cal E}$ is
given by the sum of the individual energies of the {\it decoupled}
subsystems which are described by the matrices $M_j$. Next, it is
useful to define the function
\begin{equation}
\label{fct-g}
g(p_\perp,p_1)={\rm tr} \left( B^{-1}(p_\perp,p_1) 
\partial_H B(p_\perp,p_1)\right),
\end{equation}
where the lower case symbol tr denotes the partial trace with respect
to the discrete indices $k$, $l$, cf. Eq. (\ref{matrix-B}), at fixed
$p_\perp$, $p_1$. It follows from the structure of the matrix $B_{kl}$
that $g(p_\perp,p_1+2\pi n/\lambda)=g(p_\perp,p_1)$. In addition,
choosing the height profile to be symmetric, $h(-x)=h(x)$, imposes the
condition $N_m(p_\perp,-p_1)=N_{-m}(p_\perp, p_1)$ on the matrices
$N_m$ which in turn leads after row and column permutations of the
matrix $B_{kl}$ to $g(p_\perp,-p_1)=g(p_\perp, p_1)$.  Taking at this
point the thermodynamic limit $W$, $N \to\infty$, transforms the sum
over the energies of the $N$ subsystems into an integral. Thus we
obtain, after taking the total trace over all degrees of freedom, the
final result for the Casimir force density,
\begin{equation}
\label{f-general}
F=-\frac{\hbar c}{4\pi^2}\int_0^\infty dp_\perp p_\perp 
\int_0^{\pi/\lambda} dp_1 \, g(p_\perp,p_1),
\end{equation}
where we have used the symmetries of $g(p_\perp, p_1)$. For a given
Fourier decomposition of the kernel $M$ into the matrices $N_m$, this
formula together with Eqs. (\ref{matrix-B}), (\ref{fct-g}) yields the
exact result for the force between the objects. This result has the
advantage of being particularly suited for a numerical analysis as
will be shown below. 
 
\begin{figure}[h]
  \includegraphics[width=0.95\linewidth]{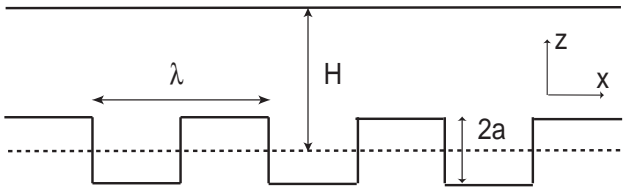}
  \vspace*{-0.3cm}
  \caption{Cross section along the $xz$-plane of the 
  two-plate configuration with translational invariance in $y$ direction.}
  \label{fig1}
\end{figure}

In the following, we apply the general formula in Eq.
(\ref{f-general}) to obtain a non-perturbative result for the force
between a flat and a rectangular corrugated surface. We choose the
height profile on a basic interval as $h(x)=a$ for $|x|<\lambda/4$ and
$h(x)=-a$ for $\lambda/4<|x|<\lambda/2$ so that $H$ is the mean
distance between the surfaces, see Fig.~\ref{fig1}. 
This profile allows for an analytical decomposition of the kernel into
matrices $N_m(p_\perp,p_1)$; details of the calculation will be
published elsewhere. The result is
\begin{eqnarray}
N_0 &=& \left(
\begin{array}{cc}
\frac{1}{2p}+\frac{e^{-2ap}-1}{4p}+\Phi_0({\bf p}) & 
\frac{e^{-pH}}{2p} \cosh(ap)\\
\frac{e^{-pH}}{2p} \cosh(ap) & \frac{1}{2p}
\end{array}\right),\nonumber\\
N_m &=& c_m \left(
\begin{array}{cc}
0 & \frac{e^{-pH}}{2p} \sinh(ap)\\
\frac{e^{-\tilde p_mH}}{2\tilde p_m} \sinh(a\tilde p_m) & 0
\end{array}\right),\nonumber
\end{eqnarray}
for odd $m\neq 0$, and
\begin{equation}
\label{matrices-N}
N_m = \left(
\begin{array}{cc}
\Phi_m({\bf p}) & 0\\
0 & 0
\end{array}\right),
\end{equation}
for even $m\neq 0$ with $\tilde p_m=\sqrt{p_\perp^2+(p_1+2\pi
  m/\lambda)^2}$, $c_m=2(-1)^{(|m|-1)/2}/\pi |m|$ and
\begin{equation}
\nonumber
\Phi_m({\bf p})=\frac{(-1)^{m/2}}{\pi^2}\!\sum_{n=-\infty}^\infty\!
\frac{f[p_\perp,p_1+(2n-1)2\pi/\lambda]}{(2n-1)(m-2n+1)},
\end{equation}
where $f(p_\perp,p_1)=(\exp(-2ap)-1)/p$.  In order to calculate the
function $g(p_\perp,p_1)$, we truncate the matrix $B_{kl}$
symmetrically around $(k,l)=(0,0)$ at order $M$ so that $k$,
$l=-(M-1)/2,\ldots,(M-1)/2$. This defines via Eq.~(\ref{fct-g}) a
series of approximations $g_M(p_\perp,p_1)$ which converges to
$g(p_\perp,p_1)$ for $M\to \infty$.

Before using the general form of the decomposition, let us consider
two instructive limiting cases. In the limit $\lambda/a \to 0$, one
would expect that $g_M(p_\perp,p_1)$ converges very rapidly since the
contributions from matrices $N_m$ decrease with $m$, cf.
Eq.~(\ref{decomp}). To test this guess, we make use of the fact that
the matrices $N_m(p_\perp,p_1+2\pi l/\lambda)$ simplify considerably
for $\lambda/a \to 0$ so that the series $g_M$ can be explicitly
obtained order by order from the truncated matrix $B_{kl}$. Using
Eq.~(\ref{fct-g}), we get
\begin{equation}
g_M(p_\perp,p_1)=\left\{
\begin{array}{cc}
-\frac{2p(1+e^{-2ap})}{1+e^{-2ap}-2e^{2(H-a)p}} & \mbox{\rm for $M=1$}\\
-\frac{2p}{1-e^{2p(H-a)}} & \mbox{\rm for $M\ge 3$}
\end{array}\right. 
\end{equation}
for positive $a$. This result indeed shows a rather rapid convergence;
the series $g_M$ is even invariant for $M \ge 3$ with increasing
dimension of the matrix $B_{kl}$. Using Eq. (\ref{f-general}), we get
the force density
\begin{equation}
\label{f-rd}
F_0 = - \frac{\pi^2}{480} \frac{\hbar c}{(H-a)^4}.
\end{equation}
Physically, this result appears to be quite natural since the most
contributing modes of the fluctuating field cannot probe the valleys
of the corrugated surface in the limit $\lambda/a \to 0$. Thus, the
surfaces experience a force which is given by the force between two
{\it flat} surfaces at reduced distance $H-a$, corresponding to Eq.
(\ref{f-rd}) \cite{Emig+01}. In the following we call this limit the
reduced distance (RD) regime.  For small $a/H$, the correction $\sim
\hbar c |a|/H^5$ to the flat surface result is {\it non-analytic} in
$a$, and thus cannot be obtained in perturbation theory.

Next, we consider the limit $\lambda/H \to \infty$.  In the limit of
small surface curvature, the pairwise summation (PWS) of renormalized
van der Waals forces is justified \cite{Derjaguin34}.  For a general
surface profile, such a summation corresponds to a spatial average
over $x$ of the Casimir energy for two flat surfaces, taken at the
local distance $H-h(x)$ \cite{Chen+02}. For the geometry of
Fig.~\ref{fig1}, this procedure obviously yields
\begin{equation}
\label{f-pws}
F_\infty = -\frac{\pi^2}{480}\frac{\hbar c}{2} 
\left(\frac{1}{(H+a)^4}+\frac{1}{(H-a)^4}\right)
\end{equation}
since 50\% of the local surface's distances are $H+a$, $H-a$, each.

\begin{figure}
  \includegraphics[width=0.90\linewidth]{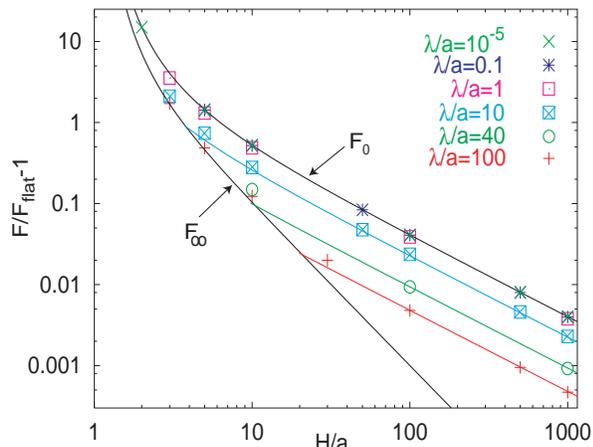}
  \vspace*{-.5cm}
  \caption{Relative increase of the Casimir force density $F$ for the
    geometry of Fig.~\ref{fig1} compared to the result $F_{\rm
      flat}=-\pi^2/480 \times \hbar c/H^4$ for two flat surfaces.}
  \label{fig2}
\end{figure}

\begin{figure}
  \includegraphics[width=0.99\linewidth]{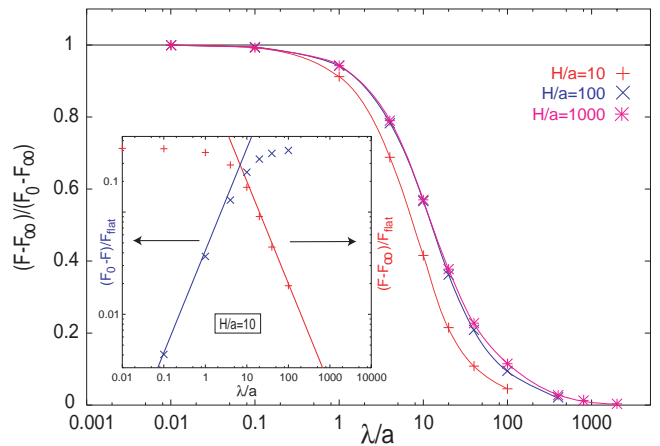}
  \vspace*{-.8cm}
  \caption{Crossover between the PWS ($F \to F_\infty$) and the RD ($F
    \to F_0$) regimes for fixed $H/a$. Inset: Scaling behavior of $F$
    for the approach to $F_0$ for $\lambda \to 0$ and to $F_\infty$
    for $\lambda \to \infty$.}
  \label{fig3}
  \vspace*{-.5cm}
\end{figure}

In order to see how the previously discussed limits fit into the
complete theory as described by Eq.~(\ref{f-general}), one has to
resort to a numerical analysis. The recipe for this analysis is
straightforward: At fixed order $M$, the truncated matrix $B_{kl}$ is
calculated from Eqs.~(\ref{matrices-N}), then its inverse is
constructed to obtain the $M^{\rm th}$ order approximation $g_M$ to
the function $g$. Finally, numerical integration in
Eq.~(\ref{f-general}) yields a corresponding series of force densities
which can be extrapolated to the final result for $M \to \infty$. The
results are summarized in Figures \ref{fig2} and \ref{fig3}. For fixed
$\lambda/a$, the relative {\it increase} $\delta =F/F_{\rm flat}-1$ of
the force compared to the force between two flat surfaces is shown in
Fig.~\ref{fig2}. At small $\lambda/a$ the exact result $F_0$ of
Eq.~(\ref{f-rd}) is recovered. For larger $\lambda/a$ there is a
crossover between two scaling regimes: For $H \gg \lambda$ the RD
scaling $\delta \sim H^{-1}$ remains valid with an amplitude which
decreases with increasing $\lambda/a$. For $H \ll \lambda$ the PWS
regime is entered with $\delta \sim H^{-2}$. As a consequence, the
force is always {\it attractive} with $F_0$ as upper and $F_\infty$ as
lower bound.  Fig.~\ref{fig3} shows the crossover between the PWS and
the RD regimes for fixed separations $H/a$. Independent of $H/a$, the
crossover appears at $\lambda/a \approx 10$. The two limits are
approached as $(F_0-F)/F_{\rm flat}\sim \lambda/a$ for $\lambda \to 0$
and $(F-F_\infty)/F_{\rm flat}\sim a/\lambda$ for $\lambda \to
\infty$, cf. the inset of Fig.~\ref{fig3}.


The reported results may stimulate new developments in a large class
of systems with fluctuation induced interactions. Most intriguing
might be the possibility of a repulsive force for the full
electromagnetic field due to anomalous contributions from the TE modes
\cite{Emig+01}. Further extensions may include many-body interactions
for more than two objects and the dynamic Casimir effect.

\begin{acknowledgments}
  We would like to thank R.~Golestanian, A.~Hanke, M.~Kardar and
  B.~Rosenow for useful discussions.  This work was supported by the
  Deutsche Forschungsgemeinschaft through the Emmy-Noether grant
  No. EM70/2-1 and by the NSF grant No. DMR-01-18213 at MIT.
\end{acknowledgments}


\begin{thebibliography}{23}
\bibitem{Hanke+02} In general, the kernel $M$ is multiplied by the
  metric functions of the surfaces. However, these factors can be
  ignored here since they drop out in the result for the force; we are
  indebted to A. Hanke for calling attention to this fact in the
  context of correlation functions, cf. A. Hanke and M.
  Kardar, Phys.  Rev. E {\bf 65}, 046121 (2002).
\end{thebibliography}
\end{document}